\documentclass[10pt, conference,a4paper]{IEEEtran}

\IEEEoverridecommandlockouts
\usepackage{cite}
\usepackage{amsmath,amssymb,amsfonts}
\usepackage{algorithmic}
\usepackage{graphicx}
\usepackage{url}
\usepackage{textcomp}
\usepackage{xcolor}
\usepackage[utf8]{inputenc}
\usepackage[T1]{fontenc}

\usepackage{fontawesome5}

\usepackage{subcaption}
\usepackage{graphicx}
\usepackage{scrextend}
\usepackage{tikz}
\usepackage{tkz-tab}
\usepackage{multirow}
\usepackage{latexsym}
\usepackage{amssymb}
\usepackage{amsmath}
\usepackage{subcaption}
\usepackage{mathtools}
\usepackage[]{moresize}
\usepackage{soul}
\usepackage{enumitem}   
\usepackage{subfiles}

\usepackage{listings}

\lstdefinelanguage{json}{
    basicstyle=\ttfamily\small,
    numbers=none,
    showstringspaces=false,
    breaklines=true,
    frame=single,
    columns=fullflexible,
    keepspaces=true,
    string=[s]{"}{"},
    comment=[l]{:},
    morecomment=[l]{,},
}

\lstdefinestyle{jsonstyle}{
    basicstyle=\ttfamily\small,
    stringstyle=\color{red},
    commentstyle=\color{green!60!black},
    keywordstyle=\color{blue},
    showstringspaces=false,
    breaklines=true,
    frame=single,
    columns=fullflexible,
    keepspaces=true,
    upquote=true
}

\usepackage{tikz}
\usepackage{pgfplots}
\usepgfplotslibrary{statistics}
\pgfplotsset{compat=1.18}
\usetikzlibrary{positioning}

\usepackage{tabularx,booktabs}
\newcolumntype{C}{>{\centering\arraybackslash}X} % centered version of "X" type
\newcolumntype{b}{X}
\newcolumntype{s}{>{\hsize=.5\hsize}X}
\newcolumntype{v}{>{\hsize=.3\hsize}X}

\def\BibTeX{{\rm B\kern-.05em{\sc i\kern-.025em b}\kern-.08em
		T\kern-.1667em\lower.7ex\hbox{E}\kern-.125emX}}
		
\usepackage[
top    = 1.9cm,
bottom = 3.90cm,
left   = 1.45cm,
right  = 1.45cm]{geometry}

\begin{document}
	
%%%%%%%%%%%%%%%%%%%%%%%%%%%%
%% METADATA
%%%%%%%%%%%%%%%%%%%%%%%%%%%%
%\title{DEMO title} 
\title{Toward Experimentation-as-a-Service in 5G/6G: The Plaza6G Prototype for AI-Assisted Trials}

\author{
        \IEEEauthorblockN{Sergio Barrachina-Mu\~noz, Marc Carrascosa-Zamacois, Horacio Bleda, Umair Riaz,\\Yasir Maqsood, Xavier Calle, Selva V\'ia, Miquel Payaró, Josep Mangues-Bafalluy}
    \IEEEauthorblockA{
        \textit{Centre Tecnològic de  Telecomunicacions de Catalunya (CTTC/CERCA)}, Barcelona, Spain \\
        \{sbarrachina, mcarrascosa, hbleda, uriaz, ymaqsood, xcalle, svia, mpayaro, jmangues\}@cttc.cat
    }
    \thanks{\textbf{This work was partially funded by Spanish MINECO grants TSI-064100-2022-16/-2023-26 (Plaza6G/Plaza6G+) and Grant PID2021-126431OB-I00 (ANEMONE) funded by MCIN/AEI/ 10.13039/501100011033 and by “ERDF A way of making Europe”}}
    \vspace{-1cm}
}

\maketitle

%\tableofcontents

% Enable page numbering
\thispagestyle{plain}
\pagestyle{plain}

%%%%%%%%%%%%%%%%%%%%%%%%%%%%
%% ABSTRACT
%%%%%%%%%%%%%%%%%%%%%%%%%%%%
\begin{abstract}

% A BIT SHORTER ABSTRACT (SB)

This paper presents Plaza6G, the first operational \textit{Experiment-as-a-Service} (ExaS) platform unifying cloud resources with next-generation wireless infrastructure. 
Developed at CTTC in Barcelona, Plaza6G integrates GPU-accelerated compute clusters, multiple 5G cores, both open-source (e.g., Free5GC) and commercial (e.g., Cumucore), programmable RANs, and physical or emulated user equipment under unified orchestration. 
In Plaza6G, the experiment design requires minimal expertise as it is expressed in natural language via a web portal or a REST API. The web portal and REST API are enhanced with a Large Language Model (LLM)-based assistant, which employs retrieval-augmented generation (RAG) for up-to-date experiment knowledge and Low-Rank Adaptation (LoRA) for continuous domain fine-tuning. 
Over-the-air (OTA) trials leverage a four-chamber anechoic facility and a dual-site outdoor 5G network operating in sub-6~GHz and mmWave bands. 
Demonstrations include automated CI/CD integration with sub-ten-minute setup and interactive OTA testing under programmable propagation conditions. 
Machine-readable experiment descriptors ensure reproducibility, while future work targets policy-aware orchestration, safety validation, and federated testbed integration toward open, reproducible wireless experimentation.

% AGREED ABSTRACT BY MC AND SB

% This paper presents Plaza6G, the first operational \textit{Experiment-as-a-Service} (ExaS) platform unifying cloud resources with next-generation wireless infrastructure.
% Developed at CTTC Barcelona, Plaza6G integrates GPU-accelerated compute clusters, multiple 5G cores (e.g., Free5GC, Open5GS, OpenAirInterface CN), programmable RANs, and physical or emulated user equipment under unified orchestration.
% A web portal and REST API enhanced with a Large Language Model (LLM)-based assistant employing retrieval-augmented generation (RAG) for accessing up-to-date experiment documentation and datasets and Low-Rank Adaptation (LoRA) for continual domain fine-tuning of orchestration dialogues enable natural-language experiment design with minimal domain expertise.
% Over-the-air (OTA) trials leverage a four-chamber anechoic facility and dual-site outdoor 5G network (sub-6~GHz, mmWave).
% We demonstrate automated Continuous Integration and Continuous Deployment (CI/CD) integration with sub-ten-minute setup times and interactive OTA experimentation under programmable propagation conditions.
% Machine-readable experiment descriptors ensure reproducibility.
% Future work targets policy-aware orchestration, safety validation, and federated testbed integration to advance open, reproducible wireless experimentation.

\end{abstract}


%\vspace{-0.5cm}
%%%%%%%%%%%%%%%%%%%%%%%%%%%%
%% INTRODUCTION
%%%%%%%%%%%%%%%%%%%%%%%%%%%%
\section{Introduction} \label{section:introduction}

% Paragraph 1: Context & Motivation
The increasing complexity of next-generation (xG) wireless networks necessitates experimentation environments that transcend simulation and theoretical modeling. Cloud-native architectures, software-defined networking, and programmable radio interfaces are fundamentally reshaping wireless service design and validation. However, integrating realistic radio conditions with scalable computing resources remains challenging. %**** REVIEW COMMENT R2.2 *** Our previous work~\cite{barrachina2025experiment} introduced \textit{Experimentation-as-a-Service} (ExaS), a paradigm enabling on-demand, reproducible, and automated experimentation by embedding xG testbeds directly into Continuous Integration and Continuous Deployment (CI/CD) pipelines. While that work established the vision and architectural principles, this paper presents the first operational realization of ExaS via the \textit{Plaza6G} platform.
Our previous work~\cite{barrachina2025experiment} built upon the \textit{Experimentation-as-a-Service} (ExaS) paradigm~\cite{edgar2017experiment,boniface2022bonfire} enabling on-demand, reproducible, and automated experimentation by embedding xG testbeds directly into Continuous Integration and Continuous Deployment (CI/CD) pipelines. While that work established the vision and architectural principles, this paper presents the first operational realization of ExaS via the \textit{Plaza6G} platform.

\begin{figure}[t!]
    \centering
    \includegraphics[width=0.46\textwidth]{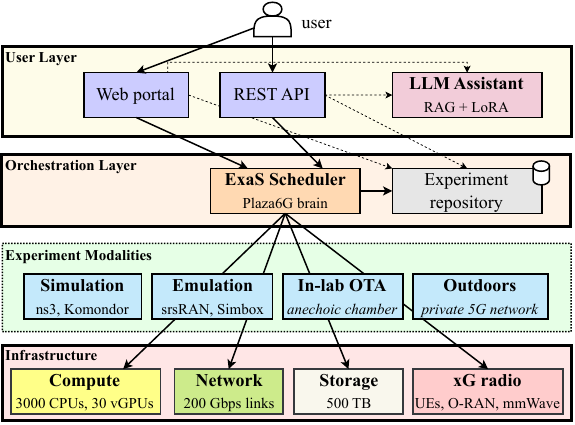}
    \caption{Conceptual architecture of Plaza6G.}
    \label{fig:plaza6g_diagram}
\end{figure}

% Paragraph 2: Plaza6G Overview
Developed at CTTC in Barcelona, \textit{Plaza6G} implements ExaS by unifying cloud computing flexibility with real 5G and 6G experimental infrastructure (Fig.~\ref{fig:plaza6g_diagram}). The platform provisions bare-metal servers, virtual machines/instances, Kubernetes clusters, GPUs, and xG radio and core networks through unified orchestration. Users interact via a web portal or API enhanced with a large-language-model (LLM)-based interface that interprets natural-language requests and automatically configures compute and wireless resources. This design enables users with minimal domain expertise to deploy complex experiments immediately, eliminating manual scripting and configuration barriers.

% Paragraph 3: Positioning & Gap
Plaza6G represents, to our knowledge, the first platform merging cloud-scale resources with real wireless infrastructure under an AI-assisted ExaS model. Existing 5G/6G testbeds typically require manual intervention, present fragmented interfaces, and demand specialized expertise, constraining accessibility and hindering reproducibility. Plaza6G addresses these limitations through integrated
automation and intelligent orchestration.

% Paragraph 4: Contribution & Paper Structure
This paper demonstrates Plaza6G's current capabilities through two representative use cases.
First, the API is used to launch in parallel three emulated 5G networks, enabling automated CI/CD network acceptance testing as part of a software developer’s pipeline, where applications under development are validated concurrently across different 5G core implementations.
Second, a real over-the-air (OTA) experiment is executed via the web portal, in which a user requests the deployment of a complete 5G network with a physical user equipment (UE) and a gNB.
These concurrent experiments highlight Plaza6G's scalability, reproducibility, and multi-tenant isolation capabilities.

%\vspace{-0.3cm}
%%%%%%%%%%%%%%%%%%%%%%%%%%%%
%% Plaza6G
%%%%%%%%%%%%%%%%%%%%%%%%%%%%
\section{The Plaza6G Platform and Resources}
\label{section:plaza6g}

\subsection{Architecture and Experiment Modalities}

Building upon the \textit{Experimentation-as-a-Service} (ExaS) model introduced in~\cite{barrachina2025experiment}, 
\textit{Plaza6G} represents its first operational instantiation—a multi-domain environment enabling automated wireless experimentation across four distinct modalities with varying fidelity levels.

The platform adopts a three-layer architecture: 
the \textit{user layer} provides web portal and REST API access augmented by an LLM-based interface interpreting intents into executable experiment graphs; 
the \textit{orchestration layer} coordinates resource allocation through policy-driven scheduling; 
and the \textit{infrastructure layer} exposes heterogeneous compute, network, and radio assets as composable services.

\begin{figure}[t!]
    \centering
    \includegraphics[width=0.48\textwidth]{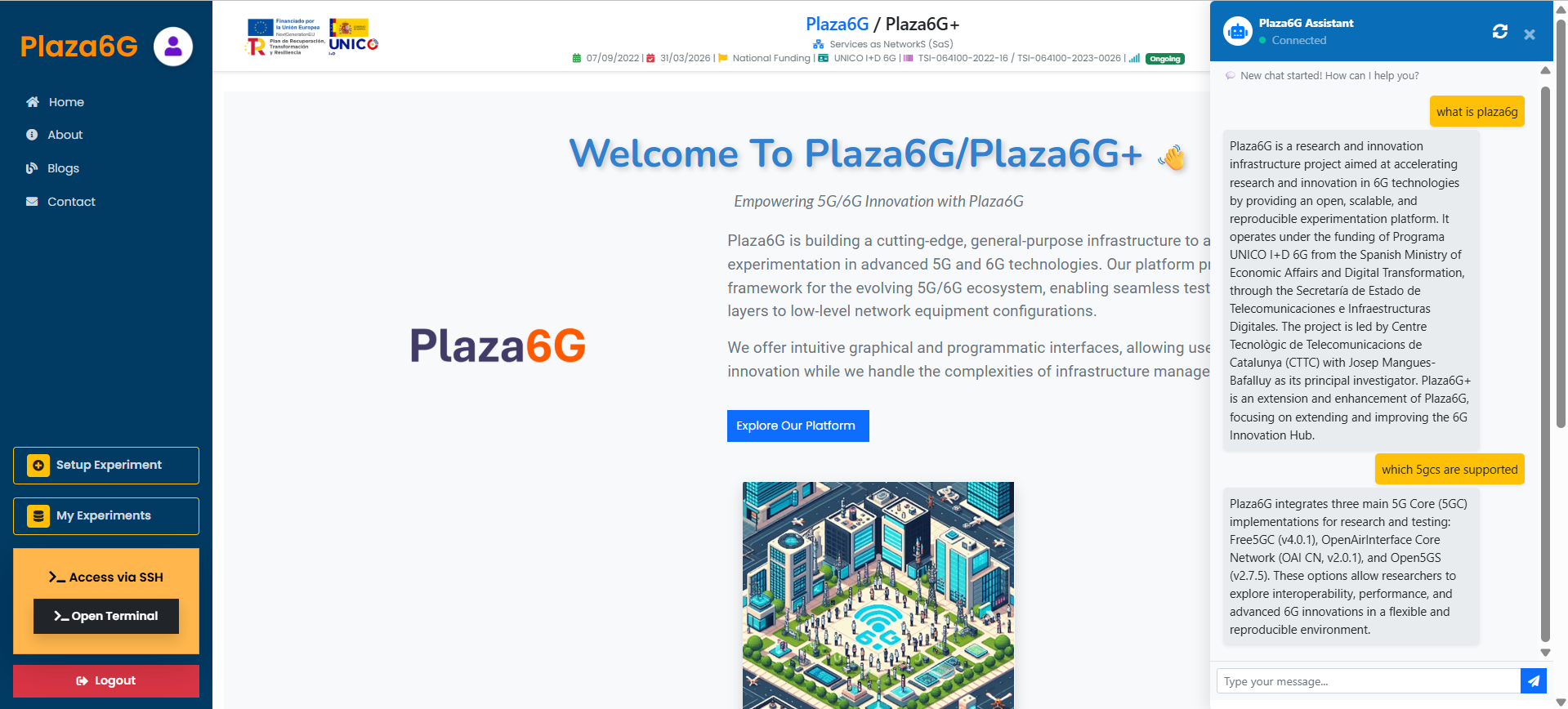}
    \caption{Plaza6G web portal with the LLM assistant.}
    \label{fig:website}
\end{figure}

User interaction combines graphical workflow composition with natural-language processing via an LLM-based assistant. 
The LLM backend, deployed locally at CTTC, leverages retrieval-augmented generation (RAG) and Low-Rank Adaptation (LoRA)~\cite{hu2022lora} to improve technical dialogue accuracy and orchestration safety while maintaining low latency. 
Fig.~\ref{fig:website} shows the portal where natural-language intents are transformed into executable workflows.

Plaza6G supports four experimentation modalities under unified orchestration:
\textbf{(i)~Simulation} employs discrete-event simulators (\textit{ns-3}~\cite{larranaga2023open}, \textit{Komondor}~\cite{barrachina2019komondor}) for large-scale protocol evaluation;
\textbf{(ii)~Emulation} instantiates virtualized protocol stacks (\textit{UERANSIM} or srsRAN UE/gNB) for rapid multi-configuration benchmarking;
\textbf{(iii)~In-lab} integrates physical equipment (e.g., \textit{Amarisoft Callbox}, commercial UEs) within a four-chamber anechoic facility for repeatable OTA validation; and
\textbf{(iv)~Outdoors} leverages a dual-site outdoor 5G network (sub-6~GHz, mmWave) for end-to-end trials under realistic conditions.
These modalities enable progressive validation from simulation through field deployment within the same framework.

\subsection{Infrastructure and Resources}

Plaza6G infrastructure spans three integrated technological domains supporting all experiment modalities. 
The \textit{compute domain} provides GPU-accelerated clusters hosting virtual machines and Kubernetes workloads with support for edge-cloud continuum deployment. 
Current capacity exceeds 3{,}000 CPU cores, 30 vGPUs (NVIDIA L40S), and approximately 500~TB of storage, enabling concurrent execution of simulation, emulation, and virtualized network function workloads. 
The \textit{network domain} offers multiple 5G core implementations such as \textit{Free5GC}, \textit{Open5GS}, or \textit{Cumucore}, supporting end-to-end network slicing and service isolation across all experiment types. 
The \textit{radio domain} comprises programmable RAN platforms including \textit{Amarisoft Callbox}, O-RAN, and srsRAN, alongside both emulated user equipment (\textit{UERANSIM}, \textit{Amarisoft Simbox}) and physical devices (commercial Android smartphones). 
Controlled OTA testing leverages the four-chamber anechoic facility ensuring repeatable radio propagation conditions for in-lab experiments, while outdoor infrastructure supports field trials as described below.

\begin{figure}[t]
    \centering
    \includegraphics[width=0.48\textwidth]{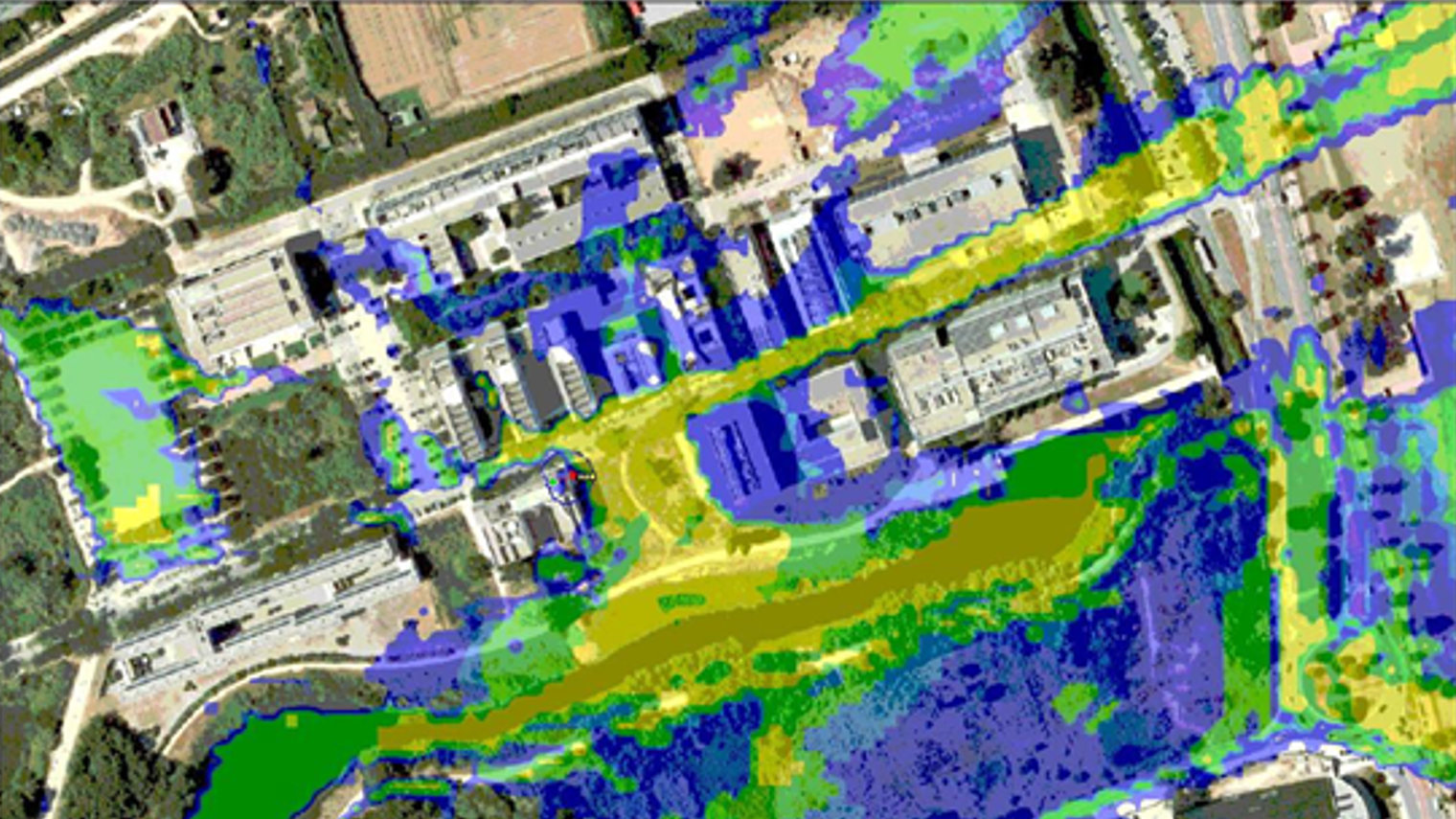}
    \caption{Coverage of the Plaza6G outdoor private 5G network at the PMT campus, operating across sub-6~GHz and mmWave bands. 
The color scale indicates received signal power, from yellow (strong) to dark blue (weak).}
    \label{fig:coverage}
\end{figure}

Plaza6G extends laboratory capabilities through a dual-site outdoor 5G network deployed across rooftop installations at the \textit{Parc Mediterrani de la Tecnologia} campus. 
Operating in both sub-6~GHz and mmWave bands, this network supports configuration with open-source or commercial 5G cores, enabling field experimentation under realistic propagation, interference, and mobility conditions. In particular, the two outdoor radio sites can also be re-configured so that they belong to two independent 5G networks to test, e.g., roaming scenarios.
Fig.~\ref{fig:coverage} presents measured coverage, demonstrating stable connectivity across the campus area served by the rooftop antenna deployment.

\subsection{Positioning Against Existing Testbeds}

Large-scale European initiatives including 5GENESIS~\cite{xylouris2021experimentation}, 5G-VINNI~\cite{gonzalez2021achieving}, 
VITAL-5G~\cite{charpentier2025utilizing}, and 5G-EVE~\cite{gupta20195g} have established comprehensive validation infrastructures for 5G technologies. 
Recent efforts such as 6G-SANDBOX~\cite{merino2024demand} target early 6G experimentation, 
while \textit{Joiner}~\cite{saunders2025joiner} federates 11 UK testbeds with automation capabilities, 
and the IEEE 5G/6G Innovation Testbed%~\cite{innovationwebsite}
~pursues end-to-end 3GPP-compliant CI/CD integration (currently under development). 
However, these platforms predominantly rely on predefined workflows requiring manual configuration, constraining both accessibility and automation potential.

\textit{Plaza6G} differentiates itself through three key innovations: 
(i)~unified orchestration spanning simulation, emulation, controlled in-lab, and field experiment modalities within a single platform, 
(ii)~AI-assisted zero-touch experimentation accessible to users without domain-specific scripting expertise, and 
(iii)~native support for the \textit{Experiment-as-Code} (ExaC) paradigm~\cite{aguilar2024experiments,edgar2017experiment,boniface2022bonfire}, 
enabling fully reproducible, version-controlled experiment definitions deployable via natural language or programmatic APIs. 
This combination of elements positions Plaza6G as a next-generation platform bridging the gap between cloud elasticity and realistic wireless experimentation across the complete validation spectrum.

%%%%%%%%%%%%%%%%%%%%%%%%%%%%
%% Use case
%%%%%%%%%%%%%%%%%%%%%%%%%%%%
\section{Demonstration Use Cases}
\label{section:usecase}

To validate Plaza6G’s operational capabilities, two representative scenarios are presented: 
(i)~automated API-driven validation integrated with external CI/CD pipelines, and 
(ii)~controlled OTA experimentation with physical 5G equipment. 
%These use cases highlight Plaza6G’s ability to embed network acceptance testing directly within software workflows while enabling reproducible OTA trials under programmable radio conditions. 
%The platform also supports field experiments over its dual-site outdoor 5G network spanning sub-6~GHz and mmWave bands, %\footnote{The outdoor private 5G network is undergoing final validation with the equipment vendor at the time of writing.} 
%confirming readiness for end-to-end multi-environment testing.
These use cases highlight Plaza6G’s ability to embed network acceptance testing within software workflows while also supporting reproducible OTA and field experiments across sub-6~GHz and mmWave environments.

\subsection{Emulated Use Case: CI/CD Network Acceptance via API}    \label{sub:emulated}

This scenario demonstrates Plaza6G as an automated validation stage within continuous integration workflows. 
A CI/CD pipeline triggers experimentation via a REST API call expressed in natural language.
For instance, to validate application performance across different 5G core implementations, a user can submit:

%\begin{minted}[breaklines]{json}
\begin{lstlisting}[language=json, style=jsonstyle]
{
  "user_request": "Deploy <my_app> across three 5G cores (Open5GS, Free5GC, OAI-CN) and verify <my_kpi> exceeds threshold for test approval."
}
\end{lstlisting}
%\end{minted}
\normalsize

%While structured JSON descriptors following API documentation are also supported, the natural-language interface significantly lowers adoption barriers by eliminating manual configuration syntax.

For the sake of simplicity, in this demonstration, \texttt{my\_app} is instantiated as \textit{iperf3} and \texttt{my\_kpi} is mean throughput with an acceptance test threshold of 50~Mbps.
The LLM backend interprets the request, identifies the application under test, target cores, and success criteria, then generates an experiment plan. 
The system returns one of three responses (\textit{approved}, \textit{clarification required}, or \textit{denied}) based on resource availability and policy constraints. 
CI/CD pipelines may proceed automatically upon approval or incorporate human-in-the-loop review, depending on organizational trust policies. %For the sake of experimentation, for the moment, automatic approval and human-in-the-loop possibilities are enabled.

Each experiment instantiates a complete emulated 5G system comprising \textit{UERANSIM}-based UE and gNB connected to one of the three 5G core implementations. 
A dedicated Data Network Name (DNN) virtual machine hosts the application server (\textit{iperf3 -s}), while the emulated UE executes \textit{iperf3 -c} for 2 minutes, generating TCP and UDP traffic for benchmarking. 
The ExaS scheduler manages the complete lifecycle (resource allocation, UE/gNB/5GC/DNN instantiation, measurement collection, and teardown) through an asynchronous scheduler. 
All three experiments execute concurrently on isolated compute pools, with comprehensive telemetry such as throughput, latency, and CPU utilization automatically archived.

\begin{figure}
    \centering
    \includegraphics[width=1\linewidth]{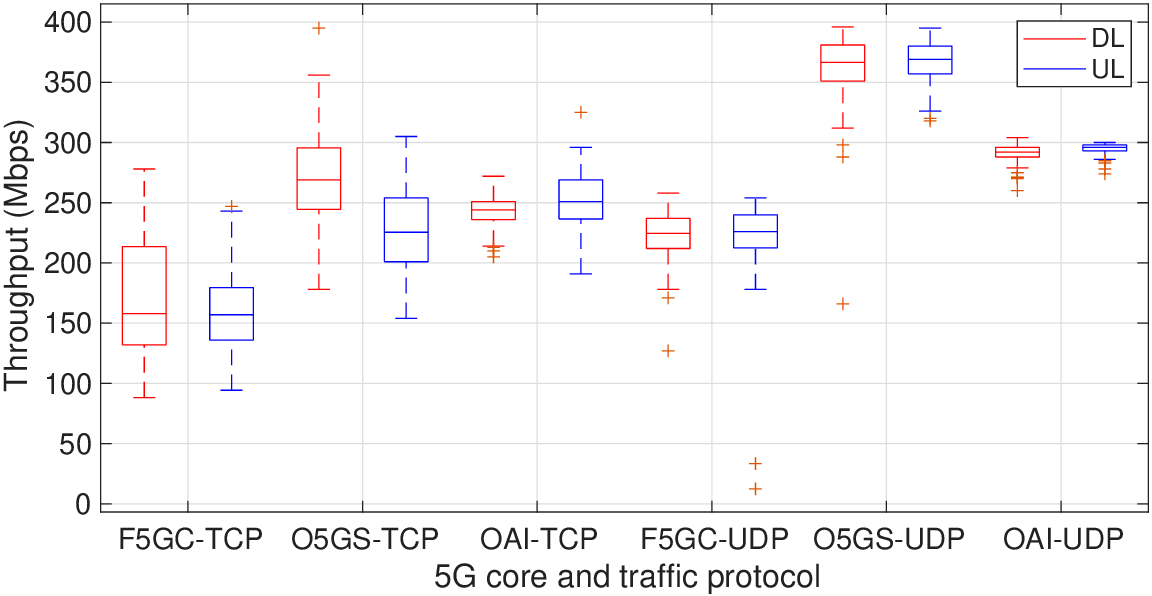}
    \caption{Throughput performance comparison for both UDP and TCP traffic across three 5G core implementations executed concurrently via Plaza6G API. Tests duration of 120 seconds, with throughput logged every second.}
    \label{fig:throughput}
\end{figure}

Figure~\ref{fig:throughput} depicts representative throughput distributions across the three 5G cores.
All measured mean values exceed the specified 50~Mbps threshold, demonstrating consistent performance for user's app under test. 
From a DevOps perspective, this enables ``network acceptance testing" where CI/CD pipelines advance to staging or production only after minimum KPI thresholds are satisfied. 
Measured setup time remains below ten minutes per experiment, reducing configuration effort by over an order of magnitude compared to manual procedures.

\subsection{In-lab use case: Over-the-Air Controlled Experiment}

The second scenario illustrates Plaza6G’s capability to automate the provisioning of complex physical experimentation environments while deliberately supporting human-in-the-loop experimentation.
Unlike the CI/CD use case we discussed before, which targets fully automated and script-driven validation, this scenario is designed for exploratory and interactive experimentation, where users manually access network elements and conduct measurements without predefined execution scripts.

\begin{figure}[t!]
    \centering
    \includegraphics[width=0.48\textwidth, trim=0 300 0 400, clip]{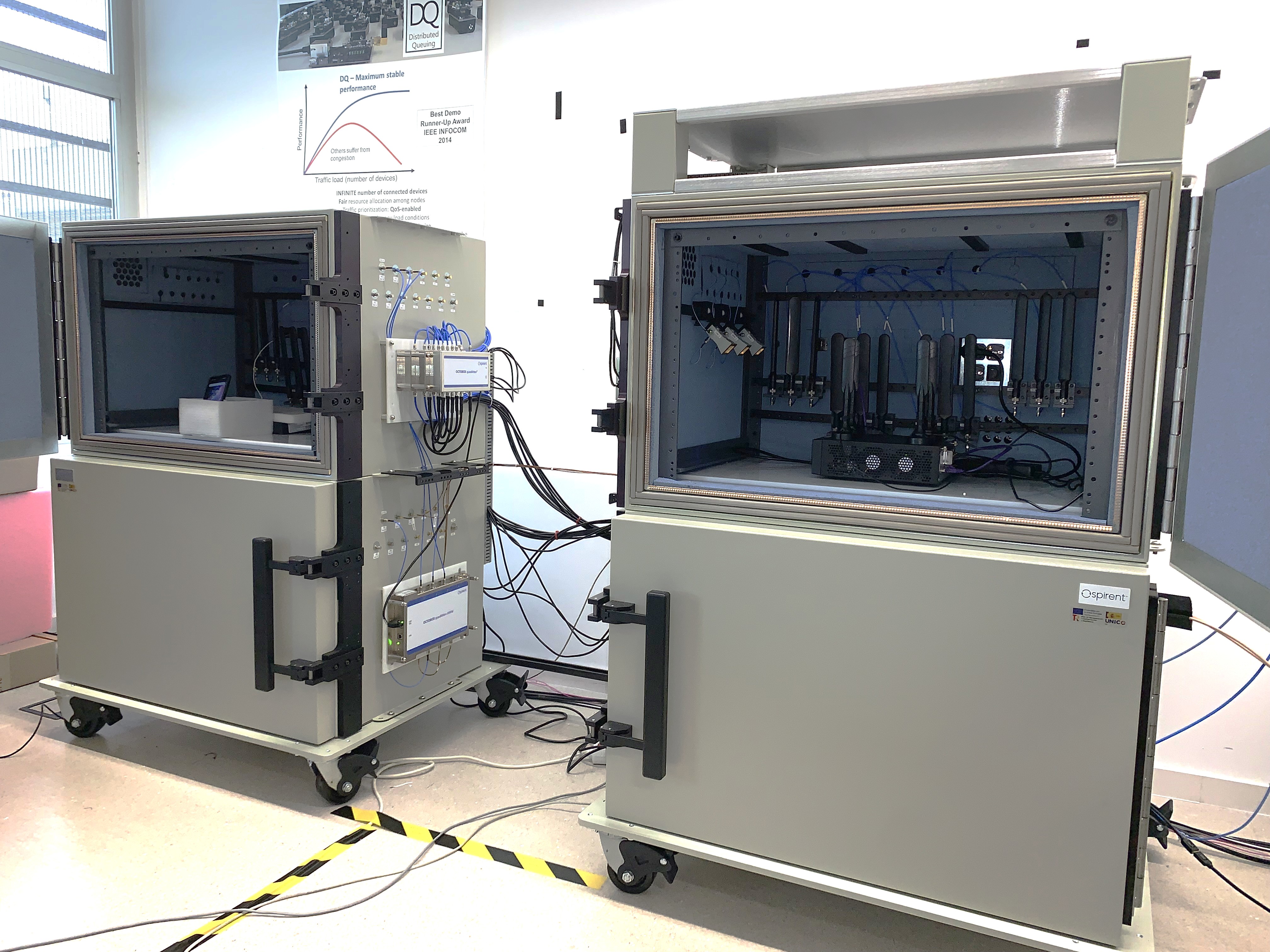}
    \caption{Anechoic chamber with an Android UE and an Amarisoft Callbox for controlled OTA experimentation.}
    \label{fig:anechoicchamber}
\end{figure}

The experiment provisioning phase is fully automated. Using the Plaza6G web portal and its LLM assistant, the user selects an experiment template that deploys a \textit{Free5GC} core, an \textit{Amarisoft Callbox} gNB, and a commercial Android smartphone acting as UE. 
Templates expose a curated set of commonly used 5G parameters (e.g., 100~MHz bandwidth, MIMO) to simplify initial configuration, while allowing users to override or refine parameters either manually or through the LLM assistant. 
The gNB and UE are placed in separate chambers of the four-chamber anechoic facility, enabling programmable control of path loss, attenuation profiles, and interference conditions. 
Figure~\ref{fig:anechoicchamber} shows the physical OTA experiment setup.
The scheduler automatically provisions compute, core, and radio resources and establishes end-to-end connectivity, after which the environment is handed over to the user for interactive experimentation.

Once the setup is complete, the user remotely accesses the UE via \textit{Vysor} to manually execute application-level tests, install software, or explore network behavior under controlled radio conditions. 
By progressively adjusting inter-chamber attenuation, users can interactively study the impact of channel degradation on throughput, latency, and perceived quality of experience. 
This mode of operation supports exploratory studies such as adaptive streaming behavior, application robustness to radio impairments, and edge–cloud service performance under varying link quality, without constraining the experiment to predefined workflows.

Throughout the session, the LLM assistant can provide contextual information on current signal conditions, active configurations, and runtime statistics in natural language, assisting users in interpreting observations while retaining full control over experiment execution.

%%%% **** REVIEW 3.4 **** Section Organization: Section III-C (LLM Assistant) describes a feature, not a use case. ---> MOVED TO CONCLUSIONS SECTION (OUTLOOK)

%\vspace{-0.1cm}
%%%%%%%%%%%%%%%%%%%%%%%%%%%%
%% Conclusions
%%%%%%%%%%%%%%%%%%%%%%%%%%%%
\section{Discussion and Outlook}
\label{section:conclusion}

%%%% **** REVIEW 3.4 **** Section Organization: Section III-C (LLM Assistant) describes a feature, not a use case. ---> MOVED FROM SECTION 3 TO CONCLUSIONS SECTION (OUTLOOK)

\subsection{Reproducibility and Workflow Traceability}

%All Plaza6G experiments follow a unified orchestration workflow in which every action, from initial resource provisioning to final teardown, is automatically logged as a machine-readable descriptor. 
%Each descriptor records software versions, network topology, hardware identifiers, and configuration parameters, ensuring that experiments can be reconstructed under identical or modified conditions. 
All Plaza6G experiments follow a unified orchestration workflow in which every action, from initial resource provisioning to final teardown, is logged as a machine-readable descriptor capturing software versions, network topology, hardware identifiers, and configuration parameters.
These descriptors are archived in a searchable experiment repository that supports version control and longitudinal performance tracking across multiple runs.

In addition to local traceability, Plaza6G maintains a consistent naming and indexing scheme that links experiment descriptors, orchestration logs, and collected metrics. 
This structured organization enhances repeatability while allowing users to audit the full lifecycle of an experiment, from creation to completion, through a unified web interface. 
To foster transparency and community reproducibility, the detailed procedure for replicating the emulated use case presented in Section~\ref{sub:emulated} has been published on the \textit{protocols.io} platform. 
This companion protocol demonstrates the same workflow executed via the Plaza6G web portal rather than the API, providing additional insight into the graphical interface and user interaction process.\footnote{The complete experimental protocol, including setup, execution, and data collection steps, is available at \url{https://www.protocols.io/view/plaza6g-experiment-reproduction-protocol-use-case-a-dm6gpm6pjgzp/v1}.}

\subsection{Conclusions and future work}

The demonstrations in Section~\ref{section:usecase} validate \textit{Plaza6G} as a practical realization of the ExaS paradigm, unifying radio, core, and compute infrastructure under automated orchestration. 
LLM-assisted interfaces and programmable APIs reduce the expertise required for wireless experimentation, enabling reproducible, concurrent trials with setup times below ten minutes. 
By coupling data-center automation with xG infrastructure, Plaza6G turns network experimentation into an on-demand cloud service for developers, researchers, and vendors.
%
%Plaza6G already provides CI/CD-compatible network acceptance testing, controlled OTA evaluation, and rapid benchmarking of 5G/6G components without manual configuration. 
%Virtualized and physical experiments coexist under a unified workflow, supporting fast iteration and consistent validation across modalities.

Future work will extend Plaza6G along several directions. 
(1)~Policy-aware orchestration will optimize scheduling for cost, energy, and radio resource usage. 
(2)~Safety and validation mechanisms will verify LLM-generated actions to ensure correctness and reproducibility. 
(3)~Planned LoRA-based fine-tuning will incrementally refine the LLM using selected orchestration logs and user dialogues, improving technical accuracy without full-model retraining. 
(4)~Federated operation with external testbeds is envisioned to enable geographically distributed, multi-domain experiments; integration with ETSI OpenSlice~\cite{etsi_openslice_website} is under study to align Plaza6G with emerging open orchestration standards.

% ----------------------------------------------------------------

\vspace{-0.1cm}
%% Bibliography
\bibliographystyle{IEEEtran}
\bibliography{bib}

\end{document}